\documentclass{jfm}

\usepackage{slashbox}
\usepackage{graphicx}
\usepackage{natbib}
\usepackage{amsmath}
\usepackage{amssymb} 
\usepackage{amsfonts}
\usepackage{bm}

\newcommand{\be}{\begin{equation}}
\newcommand{\ee}{\end{equation}}

\begin{document}

\title[The origin of oscillations of the large-scale circulation of turbulent convection]{The cause of oscillations of the large-scale circulation of turbulent Rayleigh-B{\'e}nard convection}

\author[Eric Brown and Guenter Ahlers]{Eric Brown$^1$ and Guenter Ahlers$^2$}
\affiliation{$^1$The James Franck Institute, University of Chicago, Chicago, IL 60637, \\
$^2$Department of Physics and iQCD, University of California, Santa Barbara, CA 93106}
\date{\today}

\maketitle

\begin{abstract}

In agreement with a recent experimental discovery by \cite{XZZCX09}, we also find a sloshing mode in experiments on the large-scale circulation (LSC) of turbulent Rayleigh-B{\'e}nard convection in a cylindrical sample of aspect ratio one. The sloshing mode has the same frequency as the torsional oscillation discovered by \cite{FA04}.  We show that both modes can be described by an extension of a model developed previously \cite[]{BA08a,BA08b} which consists of permitting a lateral displacement of the LSC circulation plane away from the vertical center line of the sample as well as a variation in displacements with height (such displacements had been excluded in the original model).  Pressure gradients produced by the side wall of the container on average center the plane of the LSC so that it prefers to reach its longest diameter.  If the LSC is displaced away from this diameter, the walls provide a restoring force. Turbulent fluctuations drive the LSC away from the central alignment, and combined with the restoring force they lead to oscillations.  These oscillations are advected along with the LSC.  This model predicts the correct wavenumber and phase of the oscillations, as well as estimates of the frequency, amplitude, and probability distributions of the displacements.

\end{abstract}


{\bf WORKING COPY, DO NOT DISTRIBUTE}

\section{Introduction}
Rayleigh-B{\'e}nard convection (RBC) consists of a fluid sample heated from below and cooled from above (for a review, see \cite{AGL09}).  In turbulent convection, a hot (cold) thermal boundary layer at the bottom (top) becomes unstable due to buoyancy and emits hot (cold) volumes of fluid known as ``plumes" which detach from the boundary layer and transport heat vertically.  In cylindrical containers with aspect ratio $\Gamma \equiv D/L \approx 1$ ($L$ is the height and $D$ is the diameter of the sample) these plumes contribute to the driving and are carried by a large-scale circulation (LSC).  The LSC on average forms a loop with up-flow and down-flow on opposite sides of the sample.  The dynamics of the LSC include spontaneous diffusive azimuthal meandering of the LSC orientation  \cite[]{SXX05, XZX06, BA06a, BA06b}, and re-orientations both by azimuthal {\em rotations} \cite[]{CCS97, BA06a} and {\em cessations} \cite[]{BNA05,BA06a}. All of these phenomena are now understood on the basis of a simple stochastic model \cite[]{BA07a, BA08a} derived from the Navier-Stokes equations by retaining only physically important terms, by taking volume averages that lead to two coupled ordinary differential equations, and by adding noise terms to represent the influence of the small-scale turbulent background fluctuations.  

It has long been observed that there are temperature and velocity oscillations with frequencies corresponding to the turnover times of the LSC \cite[]{HCL87, SWL89, CGHKLTWZZ89, CCL96, TSGS96, CCS97, QT00, QT01a, QT01b, NSSD01, QT02, QSTX04, FA04, SXT05,TMMS05}.    Despite these many measurements the spatial flow structure corresponding to the oscillations remained unclear for a long time.  A few years ago an analysis based on shadowgraph images first showed that the spacial structure responsible for the oscillations has a twisting, or torsional, component around the polar axis of the cylindrical container with out-of-phase oscillating azimuthal flows in the top and bottom halves of the sample \cite[]{FA04}.  The torsional shape was confirmed more recently by temperature measurements along the side wall \cite[]{FBA08}.  Very recently an additional sloshing component of the oscillating structure was discovered by \cite{XZZCX09} and described in detail by \cite{ZXZSX09}. This mode consists of a time-periodic lateral displacement of the entire LSC circulation plane away from the center line that is in phase along the entire cylinder axis. 

There have been previous attempts to find the origin of the oscillations. \cite{Vi95} presented a model that attributes the temporal oscillations to periodic emission of thermal plumes from the boundary layers.  However, recent experiments showed that there are no periodic correlations between the thermal signals of the top and bottom plates \cite[]{XZZCX09}, suggesting that the source of the observed oscillations is to be found in the bulk.  \cite{RPTDGFL06} predicted an in-phase torsional oscillation mode in ellipsoidal containers.  The oscillations only existed when the ellipticity exceeded a critical value; thus this model does not provide a mechanism for any oscillations in cylindrical containers where the oscillations are usually observed. 

The model proposed by us, which describes the spontaneous meandering, rotations, and cessations with a pair of physically motivated coupled stochastic ordinary differential equations \cite[]{BA07a, BA08a},  was extended recently \cite[]{BA08b} with perturbative terms to describe systems with various asymmetries.  In one case we predicted an in-phase torsional oscillation where the restoring force comes from the pressure gradients due to the side wall in containers with non-circular cross-sections.  At the time we did not anticipate the existence of perturbations in which the LSC plane is displaced away from the central cylinder axis, and thus we did not predict the recently characterized sloshing mode.  In the present paper we show that the model can be extended to describe the physical mechanism for and the structure of the observed oscillations of the LSC in cylindrical containers by simply relaxing the assumption of a centered LSC circulation plane.   We further show that the torsional and sloshing modes can be described simultaneously as a pair of oppositely traveling waves.  This behavior can be predicted by an extension of the model in which we remove the simplifying assumption of a planar (height-independent) LSC and by adding the appropriate advection term.

In the next section we present the results of new analyses of experimental data from \cite{BA06a, BA08a, BA08b}. First, in ~\S\,\ref{sec:method}, we discuss the sinusoidal fitting-function (SF) analysis method used by us previously and the ``temperature-extremum-extraction" (TEE) method introduced by \cite{XZZCX09}. There we also present an extended SF method which includes higher harmonics that enable the detection of the sloshing mode  and a measurement of the sloshing amplitude. In that section a comparison of time series for the slosh displacement angle obtained by the extended SF method and the TEE method  is presented. In~\S\,\ref{sec:prob} we examine the probability-distribution functions of the slosh displacement and of the torsional displacement.  In~\S\,\ref{sec:model} we extend the model presented by us previously \cite[]{BA07a, BA08a} and show that it can explain the existence of the sloshing mode as a consequence of the restoring  force that arises from a lateral displacement of the LSC circulation plane due to the turbulent background fluctuations and away from  the sample centerline. In~\S\,\ref{sec:twist} we show that this model also leads naturally to the torsional oscillation discovered several years ago \cite[]{FA04}; the origin of this mode had been unexplained before because the model had been restricted to a LSC with a circulation plane that contains the vertical sample center line. Finally, in~\S\,\ref{sec:conclusion}, we summarize the contents of this paper and discuss the broader applicability of its approach to other geometries.

\section{Experimental results for the slosh displacement}

\subsection{Analysis methods and the slosh-displacement angle}
\label{sec:method}

We use data obtained originally by \cite{BA08a} to search for the sloshing mode.  The results presented here are for a Rayleigh number $Ra \equiv \beta g \Delta T L^3 / (\kappa \nu) = 1.1\times10^{10}$ and Prandtl number $Pr \equiv \nu/\kappa = 4.38$ ($\beta$ is the isobaric thermal expansion coefficient, $g$ the acceleration of gravity, $\Delta T$ the applied temperature difference, $L$ the sample height, $\nu$ the kinematic viscosity, and $\kappa$ the thermal diffusivity).  The sample was a cylinder of aspect ratio $\Gamma \equiv D/L = 1.0$ ($D$ is the sample diameter).  The data consisted of temperature measurements by thermistors embedded in the side wall.  There were 8 thermistors, equally spaced azimuthally, at each of three different heights: the mid-height which is defined to be at $z=0$, height $z=L/4$, and height $z=-L/4$.  

\begin{figure}                                                
\centerline{\includegraphics[width=3.5in]{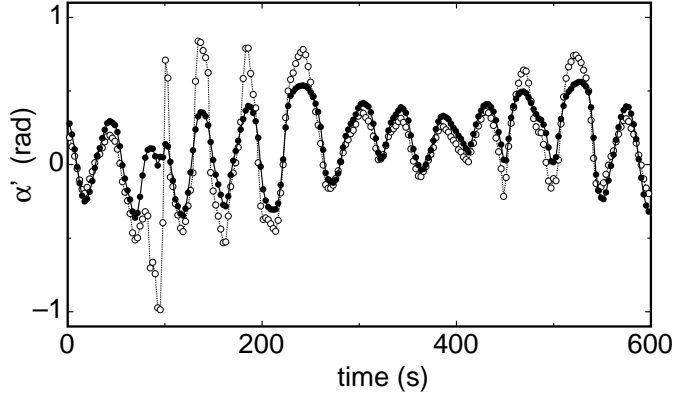}}
\caption{Time series of the slosh displacement angle $\alpha' = (\alpha_h - \alpha_c)/2$ for the middle-row thermistors at $R=1.1\times10^{10}$.  Solid circles: from the Fourier-moment method.  Open circles: from the temperature-extremum-extraction (TEE) method of \cite{ZXZSX09}. }  
\label{fig:timeseries_mid}                                       
\end{figure}

Previously, in order to determine the orientation $\theta_0$ of the LSC, we fit a sinusoidal function 

\be
T (\theta)= T_0 + \delta\cos(\theta-\theta_0)
\label{eq:temp_profile}
\ee

\noindent to the thermistor temperatures measured around the side wall at a given height \cite[]{BNA05,ABN06,BA06a,FBA08}.  This fitting function was useful for studying azimuthal diffusion and cessations of the LSC, as well as  torsional displacements and oscillations where the maximum and minimum of the temperature profile have a phase separation of $\pi$.   The sloshing oscillation was discovered and characterized \cite[]{XZZCX09,ZXZSX09} using instead the ``temperature-extremum-extraction" (TEE) method for obtaining the orientation $\alpha_h$ of the maximum and the orientation $\alpha_c$ of the minimum of the azimuthal  temperature profile.  The TEE consists of fitting a quadratic functions to  a given temperature extremum and the two  neighboring temperatures. The extremum of the quadratic fit is taken as the orientation. The sinusoidal fitting function has the advantage of using all the information from the eight temperatures at a given height, but does not allow the phase difference between $\alpha_h$ and $\alpha_c$ to differ from $\pi$. Thus it can not detect the sloshing mode. The TEE uses information from only three thermometers at a time, but does not have the phase constraint and thus can reveal the sloshing mode.  We used the TEE method to re-analyze data from earlier experiments \cite[]{BA06a, BA08a, BA08b} and found that those data also contained information about the sloshing oscillations.  We define the angles $\alpha_h$ and $\alpha_c$ with a built-in phase difference of $\pi$ so that $\alpha_h = \alpha_c$ corresponds to a centered LSC circulation plane and counter-clockwise displacements are positive.   The slosh displacement-angle is then $\alpha' \equiv (\alpha_h-\alpha_c)/2$.  An example of a time series of the slosh displacement angle $\alpha'$ obtained with the TEE method is shown as open circles in Fig.~\ref{fig:timeseries_mid}.  Since the TEE method fits to only three thermistor readings whereas the sinusoidal fit uses eight, the TEE method is more susceptible to being biased by local temperature fluctuations when the signal is very noisy [in a typical case, the root-mean-square noise amplitude due to turbulent fluctuations is approximately  $0.28\delta$ at $R=1.1\times10^{10}$  \cite[]{BA07b}].  Thus we developed a method of characterizing the side-wall temperature-profile based on Fourier moments.  The dominant term was the cosine fitting function Eq.~\ref{eq:temp_profile}.  We took the Fourier series of the difference between that fit and the actual signal.  While cosine terms are still symmetric and thus cannot contribute to a slosh displacement, the antisymmetric sine terms can characterize the off-center displacement as a perturbation of the dominant cosine term.  We calculated the sinusoidal Fourier amplitudes  

\be
A_n = [T-T_0-\delta\cos(\theta-\theta_0)]\sin[n(\theta-\theta_0)] \ .
\label{eq:fouriermoments}
\ee

\begin{figure}                                                
\centerline{\includegraphics[width=3.25in]{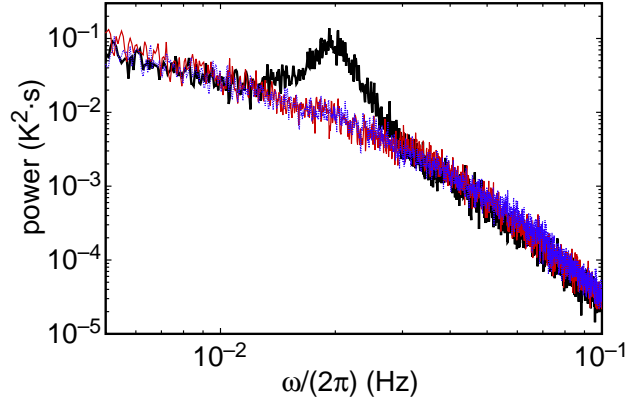}}
\caption{The power spectra of the sinusoidal Fourier moments of the temperature profile $T(\theta)$ for the middle-row thermistors at $R=1.1\times10^{10}$.  Thick solid (black online) line: 2nd order moment ($n=2$). Thin solid (red online) line: 3rd order moment ($n=3$). Dotted (blue online) line: 4th order moment ($n = 4$).  The power spectra were smoothed for clarity.}  
\label{fig:fouriermoments}                                       
\end{figure}

\noindent The $n=1$ term was found to be zero, because all it would do is shift the temperature profile, which had already been taken care of by the fit of $\theta_0$.  Higher-order terms showed strong fluctuations, and the power spectra of mode amplitudes corresponding to  $n=2,3,4$ are shown in Fig.~\ref{fig:fouriermoments}.  These higher-order moments all have the same background, similar to the backgrounds of other power spectra measured in the same system \cite[]{FBA08}.  The $n=2$ term has a peak in the power spectrum at a frequency of $0.02011\pm 0.00003$ Hz, equal to that observed by cross-correlations of thermistors on opposite sides of the sample which was 0.02008 Hz for this $R=1.1\times 10^{10}$ \cite[]{BFA07}.  This suggests that the cross-correlation also measured the slosh frequency, as suggested originally by \cite{ZXZSX09}.  Since the $A_2$ term is the only one which differs from a noise distribution, the temperature profile can be described by the cosine term plus the $n=2$ term.  For $A_2 \ne 0$, the temperature profile is tilted so that the extrema become closer together, corresponding to a slosh displacement.  The slosh displacement-angle $\alpha'$ satisfies the equation $\delta\sin\alpha' = 2A_2\cos(2\alpha')$.  A time series of $\alpha'$ comparing this fitting method with the TEE method is shown in Fig.~\ref{fig:timeseries_mid} as solid circles.  From this time series, the oscillations reported by \cite{XZZCX09} can clearly be seen.  As expected, the modified sinusoidal fitting function is found to be smoother than the TEE method because it makes use of all 8 thermistor measurements.  All cross-correlation functions between $\alpha'$ measured at any two different heights show oscillations that are in phase with each other, consistent with a sloshing oscillation. Thus our observed oscillations are in agreement with those of \cite{XZZCX09} and \cite{ZXZSX09}.

\subsection{Probability distributions of the slosh and the torsional displacements}
\label{sec:prob}

  \begin{figure}                                                
\centerline{\includegraphics[width=5in]{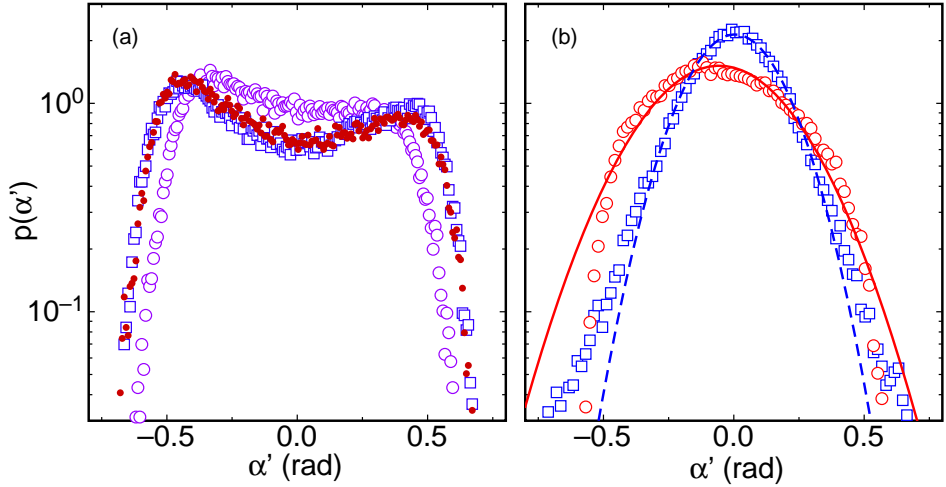}}
\caption{Probability distributions of the slosh and torsional displacement angles obtained from the Fourier-moment method at $Ra=1.1\times10^{10}$.  
(a): The slosh mode at $z = L/4$ (top row of thermistors, open squares, blue online), $z = 0$ (middle row of thermistors, open circles, purple online), and $z = -L/4$ (bottom row of thermistors, solid circles, red online).
(b): Open (red online) circles: Probability distribution of the  slosh displacement $\langle\alpha'\rangle_z$ where $\langle\alpha'\rangle_z$ is the displacement averaged over the three rows of thermistors at fixed time.  Solid (red online) line:  Gaussian fit to the peak of open circles.   Open (blue online) squares:  the probability distribution of the twist displacement angle $\theta'=[\theta_0(L/4) - \theta_0(-L/4)]/2$.    Dashed (blue online) line:  Gaussian fit to the peak of the open squares.}  
\label{fig:probsloshave}                                       
\end{figure} 
 
 The probability distributions of $\alpha'$  based on the Fourier-moment method, computed separately at each of the three vertical positions ($z = \pm L/4, 0$) for $R=1.1\times10^{10}$, are shown in Fig.~\ref{fig:probsloshave}a. They do not differ very much from each other. They are very flat and even slightly double-peaked, in agreement with the results of  \cite{ZXZSX09}.  We found that $p(\alpha')$ varied depending on the fitting method used (see ~\S\,\ref{eq:temp_profile}), which is why we developed the Fourier-moment method to filter out much of the noise and minimize the effect of local fluctuations on $\alpha'$.   We also computed the probability distribution $p(\langle\alpha'\rangle_z)$ where $\langle\alpha'\rangle_z$ is the average of $\alpha'$ over the three levels at any given moment in time. This is given in Fig.~\ref{fig:probsloshave}b as open circles.   Although on average the slosh oscillations are in phase at all three heights, the shape of the height-averaged probability distribution changes to a single-peaked distribution that is closer to Gaussian in shape. We attribute this to random fluctuations of the relative phases at the three levels which are largely uncorrelated.  A Gaussian fit to the peak gives a width of $\sigma_{\alpha'} = 0.27$ rad.   In~\S\,\ref{sec:model} we will present a model that yields such a Gaussian $p(\langle\alpha'\rangle)$.  We  will also present a model that includes the twist oscillation in~\S\,\ref{sec:twist}.  A height-averaged twist amplitude can be represented by $[\theta_0(L/4) - \theta_0(-L/4)]/2$.  This is plotted in Fig.~\ref{fig:probsloshave}b for the same data set as open squares.  A Gaussian fit to the peak gives a width of $\sigma_{\theta'} = 0.18$ rad.

 \subsection{Power spectra}

  \begin{figure}                                                
\centerline{\includegraphics[width=3.25in]{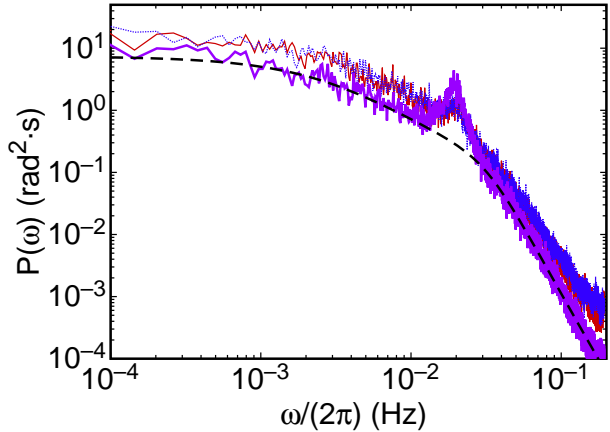}}
\caption{The power spectrum $P(\omega)$ of the slosh amplitude at $Ra=1.1\times10^{10}$. 
Thin solid line (blue online): at $z = L/4$. Thick solid line (purple online): $z = 0$. Dotted line (red online): and $z = -L/4$.  Dashed line (black online): a fit of the predicted power spectrum from Eq.~\ref{eq:power_spec} to the background of the mid-plane data.    }
\label{fig:powerspec_alpha}                                       
\end{figure} 

Power spectra $P(\omega)$ of $\alpha'$ are shown separately for the three vertical positions in Fig.~\ref{fig:powerspec_alpha}.  Despite the fact that $p(\alpha')$ is narrowest at the mid-plane, the peak of the power spectrum is larger at the mid-plane  in agreement with \cite{ZXZSX09}, showing that the oscillation amplitude is larger there.  Besides showing a peak which gives the oscillation frequency the measured power spectrum has several distinct scaling regions.  It reaches a constant in the limit of small frequency,  scales as $\omega^{-1}$ before the peak, and scales as $\omega^{-4}$ in the large frequency limit.   In ~\S\,\ref{sec:twist} we will present a model power spectrum that fits well to this data.

\section{Model for the sloshing oscillation}
\label{sec:model}
\subsection{The basic model}
The mid-plane orientation $\theta_0$ of the LSC in cylindrical containers can be described by the Langevin equation \cite[]{BA07a, BA08a}

\be
\ddot\theta_0 = -\frac{\delta\dot\theta_0}{\delta_0\tau_{\dot\theta}} +f_{\dot\theta}(t) \ .
\label{eq:lang_theta}
\ee

\noindent The damping term is due to the rotational inertia of the LSC which is represented by the measured instantaneous temperature amplitude $\delta$, the average amplitude $\delta_0$, and the damping timescale $\tau_{\dot\theta}$.   The stochastic driving term $f_{\dot\theta}(t)$ is assumed to have a Gaussian white noise distribution and is presumably due to turbulent fluctuations.  This equation describes the azimuthal meandering of the LSC orientation and,  with the addition of an equation for $\delta$, reproduces the more dramatic events of reorientations and cessations \cite[]{BA07a}.  The temperature amplitude $\delta$ fluctuates around a stable fixed-point value $\delta_0$. The details of the potential are not necessary for the purposes of understanding the slow dynamics of the LSC but become important for the discussion  for instance of cessations \cite[]{BA08b}.  The methodology used to obtain these equations involved approximate volume-averaging of the relevant terms of the Navier-Stokes equations.  This determined the important physical mechanisms behind the LSC dynamics and gave estimates of the measurable physical parameters which were typically good within about a factor of 2 \cite[]{BA07a,BA08a, BA08b}.

This model was extended by \cite{BA08b} to include perturbations that break the rotational invariance of the original model. In particular, it provided a qualitative explanation for why non-circular cross-sections cause the LSC circulation plane to align along the longest diameter.  The predicted mechanism depends on the pressure provided by the side wall. The primary component of this pressure exists also for the rotationally invariant case and provides the force to drive the LSC in a closed loop. A secondary component appears when the rotational invariance is broken and pushes the LSC towards alignment with the longest diameter. For non-circular cross-sections, the side-wall pressure yields a potential proportional to $D(\theta)^{-2}$ where $D(\theta)$ is a diameter function corresponding to the azimuthally varying container diameter for an LSC that always crosses the center of the container [\cite{BA08b}].  This gives a preferred  orientation aligned with the longest diameter with possible oscillations driven by stochastic fluctuations.

\subsection{Potential due to pressure forcing from the side walls}
 \label{sec:pressure_forcing}

\begin{figure}                                                
\centerline{\includegraphics[width=1.5in]{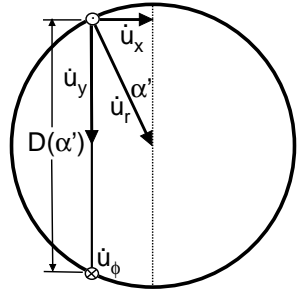}}
\caption{A horizontal cross-section showing the side wall forcing for a horizontal displacement of the LSC.  The solid vertical line corresponds to the plane of the LSC when it is displaced from a central alignment by the slosh displacement angle $\alpha'$.   The component $\dot u_y$ of the normal force $F$ provides the centripetal acceleration for the LSC and the component $\dot u_x$ acts as the restoring force for a sloshing oscillation. The primary LSC velocity $u_{\phi}$ is perpendicular to this cross-section.}  
\label{fig:sloshdiagram}                                       
\end{figure}

Here we show that the same side-wall pressure-mechanism can explain the sloshing oscillations of the LSC. The result turns out to be determined by a diameter function that depends on the off-center displacement of the LSC circulation plane. This function is analogous to the case for non-circular cross-section containers.  We will calculate model terms for containers  of height and diameter both equal to $L$ and with circular horizontal cross-sections  for comparison with experiments, but this geometrical restriction is not necessary in general (but of course the geometry has to be such that a single-loop LSC is a good approximation to the physical system). To allow for a slosh displacement we must relax the original assumption that the LSC travels through the cell center.  The diameter as a function of slosh displacement in a cylindrical container is $D(\alpha') = L\cos\alpha'$.  For simplicity we consider a planar LSC with an infinitesimal width.  We will cover a non-planar LSC in~\S\,\ref{sec:twist}.  The model can be simply explained based on a vector force-balance on the plane of the LSC at the inner edge of the viscous boundary layer at the side wall where the vertical velocity peaks.  This result can be added as a perturbation to the Langevin equation Eq.~\ref{eq:lang_theta}.  The radial pressure provided by the side wall can be broken down into two orthogonal components: one along the plane of the LSC which provides the centripetal acceleration $\dot u_y = 2u_{\phi}^2/ D(\alpha')$, and one perpendicular to the LSC plane which must then equal $\dot u_x = \dot u_y\tan\alpha'$.  A diagram of this is shown in Fig.~\ref{fig:sloshdiagram}.  The LSC velocity at the edge of the viscous boundary layer was given as $u_{\phi} = \omega_{\phi}L/2$ so that $\omega_{\phi}$ and $u_{\phi}$ independent of $\alpha'$ \cite[]{BA08b}.  Combining these terms gives the contribution to angular acceleration from pressure forces

\be
\ddot\alpha'_g =  \frac{2\dot u_x}{D(\alpha')}  = -\omega_{\phi}^2\frac{\tan\alpha'}{\cos^2\alpha'}\ . 
\label{eq:pressure_forcing}
\ee

\noindent  The pressure of the side wall provides a restoring force for the plane of LSC if it is off-center. While the LSC plane had originally been assumed to be aligned with the cell center, the off-center alignment can be generated by the action of turbulent fluctuations.  We can describe the dynamics by adding $\ddot\alpha'_g$ as a perturbative term to a dynamic equation for $\alpha'$ analogous to Eq.~\ref{eq:lang_theta}, assuming the stochastic and damping terms are the same for this off-center perturbation of the LSC as for other modes:

\be
\ddot\alpha' = -\frac{\delta\dot\alpha'}{\delta_0\tau_{\dot\theta}} - \omega_{\phi}^2\frac{\tan\alpha'}{\cos^2\alpha'} +f_{\dot\theta}(t)\ .
\label{eq:lang_alpha}
\ee

\noindent The slosh dynamics can be expressed in terms of diffusion in an azimuthal potential given by

\be
V_g \equiv -\int \ddot\alpha'_g d\alpha' = \frac{\omega_{\phi}^2 L^2}{2D(\alpha')^2} \ .
\label{eq:potential_geo}
\ee

\noindent where $D(\alpha') = L\cos\alpha'$.  This is written in terms of the diameter function with the same form as the effect of the cross-section geometry \cite[]{BA08b}.  In principle the diameter function could be calculated for slosh displacements in different cross-section geometries.  A more detailed derivation based on volume-averaging the Navier-Stokes equations throughout the sample volume was shown by \cite{BA08b} and produces the same result.  

We have derived equations assuming the width of the LSC is infinitesimal for simplicity.  Accounting for a finite width of the LSC is not expected to qualitatively change results;  it may have the effect of smoothing out  the potential near the minimum, and the potential may increase dramatically at an angle smaller than $\pi/2$ due to the LSC being forced to the side of the container.

\subsection{Oscillations}	

We can obtain a linearized Langevin equation from Eq.~\ref{eq:lang_alpha} by assuming small $\alpha'$ and assuming $\delta\approx \delta_0$ which is a good approximation in experiments except during cessations \cite[]{BA08a}. This equation is given by

\be
\ddot\alpha' = -\frac{\dot\alpha'}{\tau_{\dot\theta}} - \omega_{\phi}^2\alpha' +f_{\dot\theta}(t)
\label{eq:slho}
\ee

\noindent The resonant frequency of the deterministic part of the equation is 
\be
\omega_r = \sqrt{\omega_{\phi}^2 - 1/(4\tau_{\dot\theta}^2)}\ .
\ee
This corresponds to the natural frequency $\omega_{\phi}$ when the damping term is small compared to the restoring term, and is of the same order as the frequency of the LSC turnover as observed in experiment \cite[]{XZZCX09}.  The natural frequency scales with the turnover frequency as observed because the restoring force comes from side-wall pressure which balances with the centripetal acceleration of the LSC. 



\subsection{Probability distribution}
\label{sec:slosh_prob}

The probability distribution of the orientation can be obtained from the steady-state Fokker-Planck equation in the strong-damping limit and is

\be
p(\alpha')\propto \exp\left(-\frac{V_g}{\tau_{\dot\theta}D_{\dot\theta}}\right) \ .
\label{eq:fokker_planck_pressure}
\ee

\noindent Expanding $V_g$ up to 2nd order gives a Gaussian distribution with width $\sigma_{\alpha'} = \sqrt{D_{\dot\theta}\tau_{\dot\theta}}/\omega_{\phi}$.  Based on experimental measurements of the parameters $\tau_{\dot\theta}$ and $D_{\dot\theta}$ \cite[]{BA08a} we expect a $\sigma_{\alpha'} \approx 0.22$ rad at $Ra=1.1\times10^{10}$ and a Rayleigh-number dependence proportional to  $Ra^{-0.22}$.  The predicted width of the distribution is close to the value 0.27 rad found in \S\,~\ref{sec:prob} with no significant trend with $Ra$ \cite[]{ZXZSX09}.  Since there are oscillations, a strong-damping solution is not exact, but nonetheless gives an estimate for the amplitude. 

If we consider the prediction for larger $\alpha$, we note the restoring force diverges at $\alpha' =\pi/2$, which means that the amplitude of $\alpha'$ is strongly bounded to remain less than $\pi/2$.  This is physically significant because at $\alpha' = \pm\pi/2$ the LSC would be confined to a vertical line at the side of the container.  The typical amplitude of oscillation is on the same scale, so if the potential did not turn up sharply large fluctuations could lead to a breakup of the LSC.   This specific form of breakup has not been observed.   The dropoff in the probability distribution shown in Fig.~\ref{fig:probsloshave} occurs at a smaller displacement angle of about $0.6$ rad, which would be expected due to the finite width of the LSC.

\section{The torsional oscillation}
\label{sec:twist}

\begin{figure}                                
\centerline{\includegraphics[width=5in]{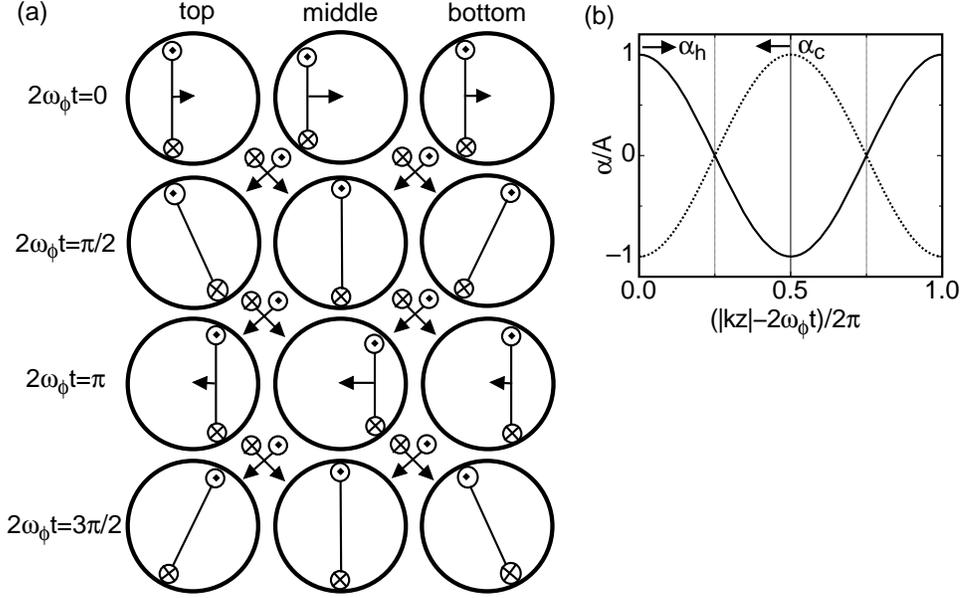}}
\caption{(a)The torsional and sloshing oscillations at different heights and phases in the Eulerian frame.  Circled dot: $\alpha_h$ where the LSC flow is upward.  Circled cross: $\alpha_c$ where the LSC flow is downward.  Connecting lines: the local plane of the LSC.  Arrows pointing horizontally from the LSC plane: restoring force on the LSC plane due to side wall pressure.   Arrows between diagrams indicate the viewing order to follow the advection of upward and downward moving plumes.  Slosh displacements are seen at $2\omega_{\phi}t = 0, \pi$ and twist displacements are seen at $2\omega_{\phi}t = \pi/2, 3\pi/2$.  (b) Displacement angles of hot and cold plumes in time; they can be seen as traveling waves moving in opposite directions.}  
\label{fig:oscdiagram}
\end{figure}

The sloshing oscillation occurs along with a torsional oscillation with out-of-phase azimuthal flows in the top and bottom halves of the sample  of approximately the same frequency and amplitude \cite[]{FA04, FBA08, ZXZSX09}.   These two modes seem to be sufficient to describe all of the observed temperature and velocity oscillations.  Even  observations that at first suggested different structures can be described as due to projections of the sloshing oscillation combined with the spontaneous meandering of the LSC. These include out-of-phase anti-correlations between temperatures at opposite sides of the sample \cite[]{ABN06,BFA07}, and apparent sloshing parallel to the LSC plane \cite[]{ZXZSX09}.  Thus if we can describe both sloshing and torsional oscillations, we can describe all of the observed oscillatory dynamics.  Some measurements indicated a difference of up to 20\% between two observed frequencies measured by different methods depending on $Ra$ \cite[]{BFA07}, but for now we will consider this as a perturbation and first try to explain why the frequencies should be the same.  Both the sloshing and torsional oscillations have been described in the Eulerian frame where they appear to have a phase difference of $\pi/2$ so that the sloshing oscillation amplitude is maximized while the twist oscillation amplitude is zero and the signal at the mid-plane follows the bottom row on the hot side but follows the top row on the cold side.   It was noted by \cite{ZXZSX09} that this direction is consistent with the direction of advection through the sample with the LSC.  If we instead consider this flow in the meridionally moving Lagrangian frame of the LSC, the motion is simply a horizontal oscillation of the plumes as they are advected along with the LSC such that the hot and cold plumes are always moving in opposing directions.   The oscillation is drawn in both the Eulerian and Lagrangian frames in Fig.~\ref{fig:oscdiagram}.   This suggests that the combination of sloshing and torsional oscillations can be reinterpreted as traveling waves consisting of oppositely moving plumes.  

We now take the step of writing down the equations for these traveling waves.  The nodes of the sloshing oscillation are observed to be at the top and bottom plates, and the antinode is found at the mid-plane. \cite[]{ZXZSX09} The node of the torsional oscillation is observed to be at the mid-plane \cite[]{FBA08}.  For ease of comparison to side wall measurements we will assume the path of the LSC to be square around the edge of the side wall with a length of $4L$. This differs from the assumption of a circular LSC shape in In~\S\,\ref{sec:model} where we needed to calculate a centripetal acceleration, but the calculations can be treated independently and in each case are self-consistent.  This means that there are two peaks of the oscillation per cycle of the LSC, the wavenumber is $k=\pi/L$,  and the oscillation frequency is $2\omega_{\phi}$ if the traveling waves are moving at the speed of the LSC.  It has been measured that temperature and velocity oscillations have a frequency $U_{max}/2L$  where $U_{max}$ is the maximum of the vertical velocity profile \cite[]{QT01b}.  Particle-image velocimetry measurements have shown that $U_{max}$ overestimates the average LSC speed by about a factor of 1.2, {\it i.e.} $U_{max} \approx 1.2\langle U\rangle$ \cite[]{SXT05}.  The path length $l$ of the circulation is given by $l = 3.4L$ for large $Ra$ \cite[]{SX05}, so the oscillation frequency is $2.0 \omega_{\phi}$, which is  self-consistent with the picture of two wavelengths per cycle.  Adjusting the path length from the simplification of $4L$ to the measured value of $3.4L$ would require a small adjustment of $k$ but does not shift the frequency.  We can now write down the empirical equations of motion. For the oscillating peak positions of the hot upward-traveling waves we have


\be
\alpha_h = A\cos(\pi z/L-2\omega_{\phi} t)
\ee

\noindent and for the cold downward traveling waves we have

\be
\alpha_c = -A\cos(-\pi z/L-2\omega_{\phi} t) \ .
\ee

\noindent The  slosh displacement is

\be
\alpha' \equiv \frac{\alpha_h-\alpha_c}{2} = A\cos{\pi z/L}\cos(2\omega_{\phi}t)
\label{eq:travelingwave_alpha}
\ee

\noindent and the  twist displacement is

\be
\theta' \equiv  \frac{\alpha_h+\alpha_c}{2} = -A\sin{\pi z/L}\sin(2\omega_{\phi}t)\ .
\label{eq:travelingwave_theta}
\ee

\noindent  This form gives a sloshing oscillation amplitude of $A\cos(\pi z/L)$ which is strongest at the mid-plane, qualitatively similar to experimental observations \cite[]{ZXZSX09}.  This gives an amplitude of $A\sin(\pi z/L)$ for the twist which corresponds to $A/\sqrt{2}$ at heights $\pm L/4$ where the amplitude is measured in experiments.  We measured root-mean-square displacement amplitudes $\sigma_{\alpha'}= 0.27$ rad and $\sigma_{\theta'} = 0.18$ rad at $R=1.1\times10^{10}$ (see \S\,\ref{sec:prob}), which yield a ratio of 1.5 in good agreement with the predicted ratio of $\sqrt{2}$.

\subsection{Advected oscillation model}

Since the observed oscillations can be described by a pair of oppositely traveling waves,
 we now  extend the linear oscillation model Eq.~\ref{eq:slho} to describe these traveling waves.  To describe the torsional oscillation with a stochastic differential equations we must now allow for displacements that vary with height, i.~e. a non-planar LSC.   With variation in height $z$ the advection term moving along with the LSC is no longer zero as was assumed in earlier versions of the model \cite[]{BA07a}.  Advection along with the LSC is represented by the term $\dot\alpha_a = -u\cdot\nabla \alpha \approx \mp U\partial\alpha/\partial z$ where a pathlength of $4L$ and angular frequency $\omega_{\phi}$ gives $U = 2\omega_{\phi}L/\pi$.  The advection is upward for the hot plumes and downward for the cold plumes. Taking the time derivative of this term to add it to a 2nd order equation adds spurious  stable solutions with constant $\dot\alpha\ne 0$ which we will ignore.   We rewrite Eq.~\ref{eq:slho} including this term. Separating equations for $\alpha_h$ and $\alpha_c$ yields

\be
\ddot\alpha_h(z)= -\frac{\dot\alpha_h(z)}{\tau_{\dot\theta}} - \omega_{\phi}^2\frac{\alpha_h-\alpha_c}{2} - \frac{2\omega_{\phi}L}{\pi}\frac{\partial \dot\alpha_h}{\partial z}+f_h(z,t) 
\label{eq:alphah_eom}
\ee

\noindent and

\be
\ddot\alpha_c(z) = -\frac{\dot\alpha_c(z)}{\tau_{\dot\theta}} + \omega_{\phi}^2\frac{\alpha_h-\alpha_c}{2} +\frac{2\omega_{\phi}L}{\pi}\frac{\partial\dot\alpha_c}{\partial z}+f_c(z,t) \ .
\label{eq:alphac_eom}
\ee

\noindent Possible steady-state solutions of the corresponding deterministic equations have  the form 

\be
\alpha_h = A\exp(ink_0 z - i\omega_n t)
\label{eq:alphahgeneral}
\ee

\noindent and

\be
\alpha_c = A\exp(-ink_0 z - i\omega_n t + i\psi) \ .
\label{eq:alphacgeneral}
\ee

\noindent where $k_0=\pi/(2L)$.  For the circulation to be closed around a loop so that the plume paths are continuous, the solutions must have integer values of $n$ and a relative phase difference $\psi=(n+1)\pi$ between $\alpha_h$ and $\alpha_c$.   Note that because we defined displacements to be  positive if they are counter-clockwise from a top view, a positive displacement for upflow followed around the loop becomes a negative displacement for downflow, accounting for the extra phase of $\pi$.  If we consider the $n=1$ mode with $\psi=0$, there is an in-phase torsional displacement and an out-of-phase slosh occurs at the top and bottom. These slosh and twist displacements have a relative phase of $\pi/2$.  It turns out that the path lengths of the LSC with the out-of-phase slosh displacement and the in-phase twist displacement are exactly the same as for a centered vertical LSC up to 2nd order in $\alpha$, so there is no restoring force for the $n=1$ mode.  The $n=2$ mode corresponds to the observed oscillations of Eqns.~\ref{eq:travelingwave_alpha} and \ref{eq:travelingwave_theta}.  There is a restoring force against this slosh displacement for $n=2$ because the path length is shorter than a vertical centered LSC.  More generally, the form of the restoring force is complicated and will not give a pure sinusoidal motion. To get a sinusoidal solution as an approximation we calculate the restoring displacement with a phase difference of $\pi$ rather than across a horizontal plane as $[\alpha_h(z)-\alpha_c(-z)]/2 = \alpha_h(z)[1-\exp(i\psi)]/2$.  This displacement is equal to zero for odd $n$ and equal to $\alpha_h(z)$ for even $n$.  While all even-order modes have some restoring force, the lowest-order even mode will dominate because damping will  reduce the contributions of higher order modes.  This suggests that the observed mode should be $n=2$ and $k = \pi/L$ in agreement with the empirical equations \ref{eq:travelingwave_alpha} and \ref{eq:travelingwave_theta}.  With this restoring-term approximation, the solutions have the power spectrum

\be
P(\omega) = D_{\dot\theta}\left[(\omega^2 - \omega_{\phi}^2 -n\omega\omega_{\phi})^2 + (\omega/\tau_{\dot\theta})^2\right]^{-1} \ .
\label{eq:power_spec}
\ee

\noindent In the limiting case of weak damping, the resonant frequency is $\omega_r = (1+\sqrt{2})\omega_{\phi}$.  Resonance with a  global maximum in the power spectrum occurs if $1/(\tau_{\dot\theta}\omega_{\phi}) < 2/(1+\sqrt{2})$.  Based on independent measurements of the model parameters \cite[]{BA08a} we obtain $1/(\tau_{\dot\theta}\omega_{\phi}) = 2.2$ for $R=1.1\times10^{10}$ which appears to be slightly above the resonance threshold.  However, it was argued by \cite{BA08b} that, because of the stochastic nature of the damping term $\delta(t)/(\delta_0\tau_{\delta})$ of Eq.~\ref{eq:lang_theta}, the effective damping is much less than the fixed-point value of $1/\tau_{\dot\theta}$, perhaps by an order of magnitude. Those authors found oscillations with a natural frequency weaker by about a factor of 5 than for the slosh, so by comparison the slosh restoring force should easily be strong enough to drive the predicted traveling-wave oscillations.  For any value of the damping term that allows resonance, the global resonant frequency remains between $2\omega_{\phi}$ and $\sqrt{2}\omega_{\phi}$ so is within 20\% of the estimated measured frequency of $2.0\omega_{\phi}$.  

We fit the predicted power spectrum of Eq.~\ref{eq:power_spec} to the measured power spectrum of $\alpha'$ at the mid-plane shown in Fig.~\ref{fig:powerspec_alpha}.  We fit the background  excluding the range  $\omega_r/2 < \omega < 2\omega_r$ and allow $\omega_0$, $\tau_{\dot\theta}$, $n$, and $D_{\dot\theta}$ to be fit parameters.  The fit gives $\omega_{\phi}/(2\pi) = 0.0101\pm0.0002$, $2\pi\tau_{\dot\theta} = 6.1\pm 0.1$ s, $n=1.95 \pm 0.08$, and $D_{\dot\theta} = (1.21\pm 0.01)\times10^{-4}$.   These fit values of $\tau_{\dot\theta}$ and $D_{\dot\theta}$ are within 20\% of the parameter values obtained from independent methods based on power spectra of $\theta_0$ \cite[]{BA08a} confirming our assumption that the same values are appropriate to describe the dynamics of $\alpha'$.  The fit value based on the background gives $\omega_{\phi} = 0.50\omega_r$ where $\omega_r$ is the measured peak frequency, which is consistent with our prediction for the $n=2$ mode in the strong damping limit and with observations.  In contrast,  for a linear harmonic oscillator the natural frequency is always greater than the resonant frequency.  The fact that the fit does not produce a peak is not a concern because the peak is expected to be enhanced by fluctuations of $\delta$.  The enhancement is all near the natural frequency since $\omega_r\tau_{\dot\theta} = 0.9$.  The region with a scaling of $\omega^{-1}$ is not expected in typical linear harmonic oscillator equation and appears in the model as a result of the advective term.  The crossover to a $\omega^{-1}$ scaling with a fit value consistent with $n=2$ provides support for the advective term in the model.  

 


The probability distributions of angular displacements $p(\alpha')$ and $p(\theta')$ are predicted to be Gaussian, see Eq.~\ref{eq:fokker_planck_pressure}.  However, the displacement distribution $p(\alpha')$ at a single height was observed to be flat or even slightly double-peaked as shown by Fig.~\ref{fig:probsloshave}a.  A twist displacement for $n=2$ has a shorter path length than a centered, planar LSC and thus corresponds to a lower potential.  Since this effect is maximized at $\pi/2$ rad out of phase with the  slosh displacement, the twist may serve to further drive the oscillation to larger amplitudes, increasing the effective restoring force and giving a flatter or double-peaked probability distribution.  An average of the slosh displacements over different heights averages over different phases of the oscillation and thus will tend to give a smoother distribution.  The flat probability distribution cannot be seen in the twist because there it is only measured based on a difference in angle at different heights.  In any case, the probability distributions based on the height-averaged displacements were shown to be in good agreement with the model.  It was observed by \cite{FBA08} that the torsional oscillation amplitude tends to be larger when the LSC strength $\delta$ is weaker.  This is expected to be the case both because the restoring and damping frequencies are  physically dependent on the LSC strength and instantaneously should be proportional to $\delta$, which would tend to widen the probability distribution calculated from Eq.~\ref{eq:fokker_planck_pressure} when $\delta$ decreases.  A simple substitution does not give an exact result but we can make a qualitative prediction that the amplitude of oscillations increases as $\delta$ decreases.
 



\subsection{Containers with non-circular cross-section}

We can apply this model to other geometries by generalizing the diameter function $D(\theta,\alpha')$.  This is valid because the potential in Eq.~\ref{eq:potential_geo} has the same form as that found for non-circular containers where the diameter function $D(\theta)$ corresponds to the length of a chord through the center of a horizontal cross-section as a function of orientation $\theta$ \cite[]{BA08b}.  Since the potential scales as $D(\theta)^{-2}$ the potential minimum corresponds to a longest diameter.  In highly elliptical or rectangular cross-section containers the side wall shape was predicted to induce an in-phase torsional oscillation that corresponds to the $n=1$ mode because the restoring force is proportional to $\theta_0$ measured relative to the longest diameter \cite[]{BA08b}.  In the interest of studying oscillations in non-circular containers we calculated the generalized diameter function $D(\theta,\alpha')$ for some geometries. 




\cite{BA08b} considered containers with elliptical cross-section defined by $D(\theta) = L[1+\varepsilon\cos(2\theta)]$ with ellipticity $\varepsilon$.    An elliptical container can be considered a perturbation on a cylindrical container where  $\varepsilon$ is the small parameter.    A second order expansion around the longest diameter gives $D(\theta, \alpha') = L(1-2\varepsilon\theta^2 - \alpha'^2/2)$.    The $n=2$ mode should be dominant for small $\varepsilon$ while the $n=1$ mode is overdamped, but once $\varepsilon$ is large enough that the $n=1$ mode is underdamped it should soon become dominant because it occurs at lower frequency so the peak of its power spectrum is reduced least by damping.  The restoring forces for the $n=1$ and $n=2$ modes are equal for $\varepsilon = 1/4$ so at that point we would expect the $n=1$ mode to dominate but because of the complicated effect of fluctuations of $\delta$ it is difficult to give a more precise value for a transition.  

In a rectangular container of horizontal aspect ratio $\Lambda$ the potential becomes asymmetric around the a diagonal.  A first order expansion around a diagonal gives the diameter function $D_{\pm} \approx [1-\Lambda^{\pm1}(|\theta| +|\alpha'|)]D_m$ where the $\pm$ terms correspond to the two sides of the longest diameter $D_m$.  The scaling with $|\alpha'|$ rather than $\alpha'^2$ should give a non-sinusoidal oscillation and changes the expected $p(\alpha')$ to a double-sided exponential as explained by \cite{BA08b} for the in-phase torsional oscillation.  The slosh and twist displacements provide equal contributions to the restoring force so we expect the $n=1$ mode to dominate in rectangular containers because it is at a lower frequency.

We note that the tilt-induced oscillations studied by \cite{BA08b} are similar to the predicted cross-section dependent oscillations because the restoring force is proportional to $\theta'$, so this restoring force can drive the $n=1$ mode.   The oscillations found by \cite{BA08b} had frequencies smaller than $\omega_{\phi}$ when induced by a small tilt angle and did not have an advective scaling regime in the power spectrum with $P(\omega)\propto \omega^{-1}$.   Thus they were best described without the advective term in the model.  This can be the case because the restoring force is relative to fixed locations in the container and can perturb the LSC without the requirement of producing closed oscillations.  We presume that either type of mode can exist for cross-section dependent oscillations where the non-advective mode occurs for small asymmetries and the $n=1$ advective mode only occurring for larger asymmetries.

\section{Conclusions}
\label{sec:conclusion}

We used a physically motivated stochastic model to explain the observed torsional and sloshing oscillations.  The proposed mechanism for the restoring force is the pressure produced by the side wall to contain the LSC which can have a component to drive the LSC towards a shape with a longer path length.   This result came from extending an earlier model used to predict that the LSC would align with the longest diameter in a container with non-circular cross section by simply relaxing the assumption that the LSC travels through the cell center.  By further relaxing the assumption of a planar LSC and adding the necessary advective term which was zero before, this model was extended to describe an advected oscillation which contains both the observed torsional and sloshing modes.  Considering pressure gradients due to the side walls, advection along the direction of the LSC, inertial damping, and a stochastic driving force as the relevant terms, we can explain the structure, frequency, and amplitude of the observed oscillations.    The apparent torsional and sloshing oscillations can now be understood as a manifestation of traveling waves which are advected along with the LSC.  Since the restoring force for both oscillation modes comes from the slosh displacement, the out-of-phase torsional mode cannot exist without a sloshing mode.  

The agreement of the oscillation frequency with twice the cell-crossing time \cite[]{QT01b} was in the past considered as a sign of the alternating emissions of plumes from the thermal boundary layers \cite[]{Vi95}.  However, it was shown recently that plumes are not emitted alternately from the thermal boundary layers \cite[]{XZZCX09}.  We have now shown that this agreement can be explained by an advected oscillation.  It is notable that we never had to invoke vertical heat transport, thermal boundary layers, or spatially or temporally periodic plume emission either as driving forces or boundary conditions.  At this point we believe we can explain the origin of all of the known dynamics of the LSC using stochastic differential equations based on approximate volume-averages of the Navier-Stokes equations.  The physics has included bouyancy, viscous damping, effective damping from rotational inertia, restoring forces from pressure gradients, turbulent fluctuations, and various perturbations from asymmetries.  Thus the LSC dynamics seem to be independent of the thermal boundary layers and plumes.  The dramatic cessations were described by a different regime of the model where fluctuations in $\delta$ are important, indicating that the two sets of dynamics are independent despite the fact that the duration of a cessation is comparable to an oscillation period.  The connection is that these timescales were both fundamentally set by advection.  

While we calculated and tested results for cells with circular cross section, this model is formulated in such a way that it can be applied to cells with other geometries.  Specifically, the diameter function $D(\theta,\alpha')$ describing a horizontal LSC cross-section is used as an input for the model.  This can in principle be applied to other container geometries as long as the flow is a single convection roll.  Since this model has been so successful at describing the dynamics with a single-roll LSC, the next logical step is to attempt to apply it to more complicated buoyancy-driven convection problems, many of which involve multiple-roll LSCs.




\end{document}